\documentclass[%
 reprint,            
 superscriptaddress, 
 amsmath,amssymb,
 aps,                
]{revtex4-2}

\usepackage{tikz}
\usetikzlibrary{fit, positioning}  
\usepackage[compat=1.1.0]{tikz-feynman} 
\usepackage{dcolumn}    
\usepackage{bm}         
\usepackage{hyperref}   
\usepackage{physics}    
\usepackage{amsmath,amssymb}
\allowdisplaybreaks[3]  
\usepackage{comment}
\usepackage{braket}
\usepackage[margin=1in]{geometry}
\usepackage{graphicx}
\DeclareGraphicsExtensions{.pdf,.png,.jpg}
\usepackage{subfig} 

\begin{document}

\title{Field Theoretic Approach to Interacting Two-Body Tunneling}

\author{YE Guo}
\email{yey@reed.edu}
\affiliation{Department of Physics, Reed College, Portland, Oregon, U.S.A.}

\date{\today} 

\begin{abstract}
Two-body tunneling problems are hard to treat analytically due to the incompatibility between tunneling and perturbation theory. The lack of classical solutions of the Euclidean Lagrangian of continuous systems further thwarts semiclassical expansions. To develop an analytic theory which provides insight on interacting two-particle tunneling, we use new results to derive the Bethe-Salpeter equation of a tunneling field theory with Yukawa coupling. We show that in the 1+1D case a closed form solution in the instantaneous positive-energy (Salpeter) regime is permitted. We then compute the scattering amplitude by perturbing on interparticle interaction and recover the Lippmann–Schwinger equation to confirm physical consistency and relevancy.
\end{abstract}

\maketitle

\section{Introduction}
Experimental advancements in the recent decade, especially in the realm of cold atom physics, have given us unprecedented ability to manipulate and measure quantum tunneling processes, even for investigating under barrier phenomena \cite{Ramos2020LarmorClock}. By contrast, theoretical treatment of many particle tunneling remains challenging due to the non-perturbative nature of tunneling. Outside of single body cases, where we may solve exactly; discrete models, where it is easier to employ conventional many body formalisms \cite{Foelling2007SecondOrder}; and the large $N$ regime, where statistical effects dominate \cite{MosheZinnJustin2003LargeN}, few analytical approaches have probed the interplay between interaction and tunneling beyond mean field approximations. The continuous two particle tunneling problem, with a Hamiltonian of the form 
\begin{align}
    \hat{H}=\hat{h}(1)+\hat{h}(2)+\alpha (W(\hat{x}_1)+W(\hat{x}_2))+\beta U(\hat{x}_1-\hat{x}_2)
\end{align}
with $\hat{h}(1),\hat{h}(2)$ being the kinetic terms for particles 1 and 2, $W$ being the static potential, $U$ being the non-local interaction, and $\alpha,\beta$ being the respective interaction strength, is a typical problem of this kind. It has broken translational symmetry on top of inseparability between $x_1$ and $x_2$. These features typically prevent exact solutions, including classical solutions in imaginary time. 

Despite these difficulties, theoretical understanding of non-perturbative two-particle tunneling remains desirable due to several reasons. It is abundant in nature~\cite{Pfutzner2012TwoProton}, vastly applied in technology such as the tunnel diode, Josephson junction, and scanning tunneling microscope~\cite{Esaki1958,Josephson1962,Binnig1982STM}, and provides a simplified model for nuclear problems such as two-proton decay, where the third body is treated as a background field~\cite{Goldansky1960,
Goldansky1960, Pfutzner2012TwoProton}. 

A field theoretic approach to tunneling has recently been formulated by \citet{Zielinski2024} which naturally encompasses many particle effects with the creation and annihilation of mesons, although the mesons themselves do not interact with the tunneling barrier. Using the same field Lagrangian, we go beyond a single particle $S$-matrix by using the resummed tunneling free propagator, obtained in the previous paper \citet{Zielinski2024b}, to construct two particle ladder diagrams. Although we derive the resummed four point correlation function recursively in the form of the Bethe-Salpeter equation, which we show to be equivalent to the class of problems of (1) in the non-relativistic (NR) limit, we delegate the specifics of a closed form solution to a second study due to mathematical complexity. Here, we provide the form of the closed form solution and follow the methodology of \citet{Zielinski2024} to perturb the interparticle interaction and evaluate the tunneling amplitudes in the $\left|\frac{\beta}{\alpha}\right|\xrightarrow{}0$ limit to derive explicit tunneling amplitudes.

\section{Formalism}
\subsection{Preliminaries}
The field Lagrangian, with $\psi$ denoting the tunneling field with mass $m$, and $\phi$ denoting the meson field with mass $\mu$ coupled to $\psi$ at strength $g$, is identical to the one in \cite{Zielinski2024}:
\begin{equation}\begin{aligned}
    \mathcal{L}=\tfrac12(\partial_\mu \psi)(\partial^\mu \psi) -\tfrac{1}{2} m^2\psi^2
- \tfrac12\,eu(x)\,\psi^2 \\+\tfrac{1}{2}(\partial_\mu\phi)(\partial^\mu\phi) - \tfrac12\mu^2\phi^2
-\tfrac{1}{2} g\,\phi\,\psi^2.\\
\end{aligned}\end{equation}
The corresponding Feynman rules are the same, which we omit here for brevity. Whilst we have used differing algebraic notation, we will use the same diagrammatical rules for the free propagators: $\psi$ propagators are dashed lines and $\phi$ (meson) propagators are solid lines. 

For $u(x)=aV_0\delta(x_3)$, where the $x_i$ is the $i^{th}$ component of $x$, the $S$-matrix of   
\begin{equation}
    \mathcal{L}_{tun}=\tfrac12(\partial_\mu \psi)(\partial^\mu \psi) -\tfrac{1}{2} m^2\psi^2
- \tfrac12\,eu(x)\,\psi^2  
\end{equation}
was resummed in closed form in~\cite{Zielinski2024b}. We use it to obtain the resummed $\psi$ propagator via a process outlined in the appendix A. The result is 
\begin{widetext}
\begin{equation}\begin{aligned}
     D_\psi(p,k)=&\delta^4(p-k)\frac{i}{p^2-m^2+i\epsilon}+-ieaV_0(2\pi)^3(1+\frac{ieaV_0}{2f(p_\mu)})^{-1}\frac{i}{p^2-m^2+i\epsilon}\delta^3(p^{\mu}-k^{\mu})\frac{i}{k^2-m^2+i\epsilon}\\
     =&\delta^4(p-k)i\Delta_\psi(p)+-ieaV_0(2\pi)^3(1+\frac{ieaV_0}{2f(p_\mu)})^{-1}i\Delta_\psi(p)\delta^3(p_{\mu}-k_{\mu})i\Delta_\psi(k).\\
\end{aligned}\end{equation}
\end{widetext}
where we denote $(p_0,p_1,p_2)=p_\mu$ so $(p_\mu)^2=(p_0)^2-(p_1)^2-(p_2)^2$, and $f(p_\mu)=\sqrt{p_{0}^{2} -p_{1}^{2} - p_{2}^{2} - m^{2}}$. Momentum is notably not conserved in the $x_3$ direction due to broken translational symmetry. Diagrammatically, we represent

\begin{equation}
D_\psi(p,k)=
\begin{tikzpicture}[baseline=(v.base)] 
  \begin{feynman}
  \node[dot]         (a) at (0,0)   {};
  \node[crossed dot, minimum size=16pt] (v) at (1.8,0) {};
  \node[dot]         (b) at (3.6,0) {};

    \diagram*{
      (a) -- [scalar, edge label=\(p\)] (v)
           -- [scalar, edge label=\(k\)] (b)
    };
  \end{feynman}
\end{tikzpicture}\,\,.
\end{equation}

\subsection{Bethe-Salpeter Equation}
We approximate the four point correlation function of (2) non perturbatively by resumming all the ladder diagrams. This approximation is justified for $|g/eaV_0|\ll1$ or for large $\mu$, or in the NR regime. Creation and annihilation of the $\psi$ tunneling particles is also forbidden, which isolates two-body tunneling characteristics that should also be observable in non relativistic systems like (1). 

Since between every instance of meson exchange, the $\psi$ field could couple to the barrier an arbitrary number of times, the barrier effect is treated exactly by inserting the tunneling propagator:

\begin{equation}
\vcenter{\hbox{
\begin{tikzpicture}
    \begin{feynman}
    \vertex [dot, minimum size=4pt] (i1);
    \vertex [dot, above of=i1] (v1) {};
    \vertex [right of = v1] (b1);
    \vertex [dot, above of = v1] (v2){};
    \vertex [right of = v2] (b2);
    \vertex [above of = v2] (o1);
    \diagram*{ 
    (i1)--[scalar](v1), 
    (v1)--[scalar](v2),
    (v2)--[scalar](o1),
    (v1)--[plain] (b1),
    (v2)--[plain] (b2)
    };
    \end{feynman}
\end{tikzpicture}}} 
\;+\;
\vcenter{\hbox{
\begin{tikzpicture}[scale=0.75, transform shape, node distance=1cm]
    \begin{feynman}
    \vertex [dot, minimum size=4pt] (i1);
    \vertex [dot, above of=i1] (v1){};
    \vertex [right of = v1] (b1);
    \vertex [above of = v1] (v2);
    \vertex [crossed dot, right of = v2] (b2){};
    \vertex [dot, above of = v2] (v3) {};
    \vertex [right of = v3] (b3) {};
    \vertex [above of = v3] (o1);
    \diagram*{ 
    (i1)--[scalar](v1), 
    (v1)--[scalar](v2),
    (v2)--[scalar](v3),
    (v3)--[scalar](o1),
    (v1)--[plain] (b1),
    (v2)--[boson] (b2),
    (v3)--[plain] (b3),
    };
    \end{feynman}
\end{tikzpicture}}} 
\;+\;
\vcenter{\hbox{
\begin{tikzpicture}[scale=0.6, transform shape, node distance=1cm]
    \begin{feynman}
    \vertex (i1);
    \vertex [dot, above of=i1] (v1){};
    \vertex [right of = v1] (b1);
    \vertex [above of = v1] (v2);
    \vertex [crossed dot, right of = v2] (b2){};
    \vertex [above of = v2] (v3);
    \vertex [crossed dot, right of = v3] (b3) {};
    \vertex [dot, above of = v3] (v4) {};
    \vertex [right of = v4] (b4) ;
    \vertex [above of = v4] (o1);
    \diagram*{ 
    (i1)--[scalar](v1), 
    (v1)--[scalar](v2),
    (v2)--[scalar](v3),
    (v3)--[scalar](v4),
    (v4)--[scalar](o1),
    (v1)--[plain] (b1),
    (v2)--[boson] (b2),
    (v3)--[boson] (b3),
    (v4)--[plain] (b4)
    };
    \end{feynman}
\end{tikzpicture}}} 
\;+\;\cdots
=
\vcenter{\hbox{
\begin{tikzpicture}[scale=0.8, transform shape, node distance=1cm] 
  \begin{feynman}
    \vertex (i1);
    \vertex [dot, above of=i1] (v1){};
    \vertex [right of = v1] (b1);
    \vertex [crossed dot, minimum size=16pt, above of = v1] (v2) {};
    \vertex [dot, above of=v2] (v3){};
    \vertex [right of = v3] (b3);
    \vertex [above of = v3] (o1);
    \diagram*{
      (i1) -- [scalar] (v1),
      (v1) -- [scalar] (v2),
      (v2) -- [scalar] (v3),
      (v3) -- [scalar] (o1),
      (v1) -- [plain] (b1),
      (v3) -- [plain] (b3)};
  \end{feynman}
\end{tikzpicture}}}
\end{equation}

The sum of amputated ladder diagrams, including the free term, which we denote $L$, could then be written as
\begin{equation}
    \vcenter{\hbox{
\begin{tikzpicture}[scale=0.6, transform shape, node distance=1cm]
    \begin{feynman}
    \vertex [dot] (i1) {};
    \vertex [dot, right of = i1] (I1) {};
    \vertex [above of = i1] (v1);
    \vertex [above of = I1] (V1);
    \vertex [above of = v1] (v2);
    \vertex [above of = V1] (V2);
    \vertex [dot, above of = v2] (v3){};
    \vertex [dot, above of = V2] (V3){};
    \diagram*{ 
    (i1)--[scalar](v1)--[scalar](v2)--[scalar](v3),
    (I1)--[scalar](V1)--[scalar](V2)--[scalar](V3)
    };
    \node[draw, rectangle, thick, fill=white,
          fit=(v1)(V1)(v2)(V2),
          inner sep=4pt,
          label=center:\Large$L$] {};
    \end{feynman}
\end{tikzpicture}}} 
\:=\:
\vcenter{\hbox{
\begin{tikzpicture}[scale=0.6, transform shape, node distance=1cm]
    \begin{feynman}
    \vertex [dot] (i1) {};
    \vertex [dot, right of = i1] (I1) {};
    \vertex [crossed dot, minimum size=16pt, above of = i1] (v1){};
    \vertex [crossed dot, minimum size=16pt, above of = I1] (V1){};
    \vertex [dot, above of = v1] (v2){};
    \vertex [dot,above of = V1] (V2){};
    \diagram*{ 
    (i1)--[scalar](v1)--[scalar](v2),
    (I1)--[scalar](V1)--[scalar](V2)
    };
    \end{feynman}
\end{tikzpicture}}} 
\;+\;
\vcenter{\hbox{
\begin{tikzpicture}[scale=0.6, transform shape, node distance=1cm]
    \begin{feynman}
    \vertex [dot] (i1) {};
    \vertex [dot, right of = i1] (I1) {};
    \vertex [crossed dot, minimum size=16pt, above of = i1] (v1){};
    \vertex [crossed dot, minimum size=16pt, above of = I1] (V1){};
    \vertex [dot, above of = v1] (v2){};
    \vertex [dot,above of = V1] (V2){};
    \vertex [crossed dot, minimum size=16pt, above of = v2] (v3){};
    \vertex [crossed dot, minimum size=16pt, above of = V2] (V3){};
    \vertex [dot, above of = v3] (v4){};
    \vertex [dot,above of = V3] (V4){};
    \diagram*{ 
    (i1)--[scalar](v1)--[scalar](v2)--[scalar](v3)--[scalar](v4),
    (I1)--[scalar](V1)--[scalar](V2)--[scalar](V3)--[scalar](V4),
    (v2)--[plain](V2)
    };
    \end{feynman}
\end{tikzpicture}}} 
\;+\;
\vcenter{\hbox{
\begin{tikzpicture}[scale=0.6, transform shape, node distance=1cm]
    \begin{feynman}
    \vertex [dot] (i1) {};
    \vertex [dot, right of = i1] (I1) {};
    \vertex [crossed dot, minimum size=16pt, above of = i1] (v1){};
    \vertex [crossed dot, minimum size=16pt, above of = I1] (V1){};
    \vertex [dot, above of = v1] (v2){};
    \vertex [dot,above of = V1] (V2){};
    \vertex [crossed dot, minimum size=16pt, above of = v2] (v3){};
    \vertex [crossed dot, minimum size=16pt, above of = V2] (V3){};
    \vertex [dot, above of = v3] (v4){};
    \vertex [dot,above of = V3] (V4){};
    \vertex [crossed dot, minimum size=16pt, above of = v4] (v5){};
    \vertex [crossed dot, minimum size=16pt, above of = V4] (V5){};
    \vertex [dot, above of = v5] (v6){};
    \vertex [dot,above of = V5] (V6){};
    \diagram*{ 
    (i1)--[scalar](v1)--[scalar](v2)--[scalar](v3)--[scalar](v4)--[scalar](v5)--[scalar](v6),
    (I1)--[scalar](V1)--[scalar](V2)--[scalar](V3)--[scalar](V4)--[scalar](V5)--[scalar](V6),
    (v2)--[plain](V2),
    (v4)--[plain](V4)
    };
    \end{feynman}
\end{tikzpicture}}} 
\;+\;\cdots,
\end{equation}
or recursively:
\begin{equation}
    \vcenter{\hbox{
\begin{tikzpicture}[scale=0.6, transform shape, node distance=1cm]
    \begin{feynman}
    \vertex [dot] (i1) {};
    \vertex [dot, right of = i1] (I1) {};
    \vertex [above of = i1] (v1);
    \vertex [above of = I1] (V1);
    \vertex [above of = v1] (v2);
    \vertex [above of = V1] (V2);
    \vertex [dot, above of = v2] (v3){};
    \vertex [dot, above of = V2] (V3){};
    \diagram*{ 
    (i1)--[scalar, edge label = \(u^a\)](v1)--[scalar](v2)--[scalar, edge label = \(p^a\)](v3),
    (I1)--[scalar, edge label = \(u^b\)](V1)--[scalar](V2)--[scalar, edge label = \(p^b\)](V3)
    };
    \node[draw, rectangle, thick, fill=white,
          fit=(v1)(V1)(v2)(V2),
          inner sep=4pt,
          label=center:\Large$L$] {};
    \end{feynman}
\end{tikzpicture}}} 
\:=\:\vcenter{\hbox{
\begin{tikzpicture}[scale=0.6, transform shape, node distance=1cm]
    \begin{feynman}
    \vertex [dot] (i1) {};
    \vertex [dot, right of = i1] (I1) {};
    \vertex [crossed dot, minimum size=16pt, above of = i1] (v1){};
    \vertex [crossed dot, minimum size=16pt, above of = I1] (V1){};
    \vertex [dot, above of = v1] (v2){};
    \vertex [dot,above of = V1] (V2){};
    \diagram*{ 
    (i1)--[scalar, edge label = \(u^a\)](v1)--[scalar,edge label = \(p^a\)](v2),
    (I1)--[scalar, edge label = \(u^b\)](V1)--[scalar, edge label = \(p^b\)](V2)
    };
    \end{feynman}
\end{tikzpicture}}} 
\;+\;
\vcenter{\hbox{
\begin{tikzpicture}[scale=0.6, transform shape, node distance=1cm]
    \begin{feynman}
    \vertex [dot] (i1) {};
    \vertex [dot, right of = i1] (I1) {};
    \vertex [above of = i1] (v1);
    \vertex [above of = I1] (V1);
    \vertex [above of = v1] (v2);
    \vertex [above of = V1] (V2);
    \vertex [dot, above of = v2] (v3){};
    \vertex [dot, above of = V2] (V3){};
    \vertex [crossed dot, minimum size=16pt, above of = v3] (v4) {};
    \vertex [crossed dot, minimum size=16pt, above of = V3] (V4) {};
    \vertex [dot, above of = v4] (v5) {};
    \vertex [dot, above of = V4] (V5) {};
    \diagram*{ 
    (i1)--[scalar, edge label = \(u^a\)](v1)--[scalar](v2)--[scalar, edge label = \(v^a\)](v3)--[scalar, edge label = \(k^a\)](v4)--[scalar, edge label = \(p^a\)](v5),
    (I1)--[scalar, edge label = \(u^b\)](V1)--[scalar](V2)--[scalar, edge label = \(v^b\)](V3)--[scalar, edge label = \(k^b\)](V4)--[scalar, edge label = \(p^b\)](V5),
    (v3)--[plain, edge label = \(q\)](V3)
    };
    \node[draw, rectangle, thick, fill=white,
          fit=(v1)(V1)(v2)(V2),
          inner sep=4pt,
          label=center:\Large$L$] {};
    \end{feynman}
\end{tikzpicture}}} \;.
\end{equation}
Here we have labeled the momentum variables used for both the $\phi$ and $\psi$ field. This corresponds to the algebraic equation of
\begin{widetext}
\begin{equation}\begin{aligned}
L(p^a,p^b;u^a,u^b)=&D_\psi(p^a,u^a)D_\psi(p^b,u^b)+g^2\int D_\psi(p^a,k^a)D_\psi(p^b,k^b)D_{\phi}(q) L(v^a,v^b;u^a,u^b)\\
   & \times\delta^2(k^a+q-v^a)\delta^2(k^b-q-v^b) d^4k^a d^4k^bd^4qd^4v^ad^4v^b,
\end{aligned}\end{equation}
\end{widetext}
where $D_{\psi}$ is given by (4), and $D_{\phi}(q)=\tfrac{i}{q^2-\mu^2+i\epsilon}$. 

(9) can be drastically simplified by collapsing the $q$, $k^a,k^b$, and $v^a,v^b$ integrals. To do this we proceed to the total and relative momentum coordinates. For all three variables:
\nopagebreak[4]
\vspace{-0.5\baselineskip}
\begin{equation}
\begin{aligned}
    \xi^a+\xi^b&:=\Xi\\
    \xi^a-\xi^b&:=\xi\\
    d^4 \xi^a d^4\xi^b& \rightarrow d^4\Xi d^4\xi .
\end{aligned}
\end{equation}
We also restrict to the case where $\mathbf{P}=0$, hence $P=(E,\mathbf{0})$. Physically, this implies we treat only outgoing wave packets supported where $p^a_1=-p^b_1$, hence reducing the degrees of freedom by 3. Notably, this places no restriction on $U$ since translational symmetry is broken by the barrier, hence we are still able to describe tunneling processes of arbitrary incoming momenta, albeit the scattered particles must be traveling in opposite directions. This does not preclude two-body binding; bound states are naturally characterized and often computed in the $\mathbf{P}=\mathbf{0}$ sector, where the bound-state pole occurs at $E=E_B(\mathbf{0})$.  Substantial calculations follow, but most crucially we are able to separate the tunneling propagator legs in (8) from the kernel variables $V,v$ and show that $k^a,k^b\rightarrow[p_\mu^a,f(p_\mu^a)],[p_\mu^b,-f(p_\mu^b)]$, which implies that $k^a,k^b$ collapses completely into $p^a,p^b$ if we grant the latter to be on shell. After the process which we outline in Appendix B, we obtain
\begin{widetext}
\begin{equation}
\begin{aligned}
    L(P,p;U,u)=&D_\psi\left(\frac{P+p}{2},\frac{U+u}{2}\right)D_\psi\left(\frac{P-p}{2},\frac{U-u}{2}\right)\\
    +&\Gamma\left(\frac{P+p}{2}\right)\Gamma\left(\frac{P-p}{2}\right)g^2\int \frac{i}{(v-p')^2/4-\mu^2+i\epsilon}L(P',v;U,u)d^4v,
\end{aligned}
\end{equation}
\end{widetext}
where we have denoted for conciseness $[p_\mu^a,f(p_\mu^a)]-[p_\mu^b,-f(p_\mu^b)]:=p'$, $[p_\mu^a,f(p_\mu^a)]+[p_\mu^b,-f(p_\mu^b)]=[0,f(p_\mu^a)-f(p_\mu^b)]:=P'$, and $\Gamma$ is
\begin{widetext}
\begin{equation}\begin{aligned}
    \Gamma\left(\frac{\Xi\pm \xi}{2}\right)=\left(1\pm\frac{F([\Xi_\mu\pm\xi_\mu]/2)\pi}{f([\Xi_\mu\pm\xi_\mu]/2)}\right)\frac{i}{(\Xi\pm \xi)^2/4-m^2+i\epsilon}.
\end{aligned}\end{equation}
\end{widetext}
with 
\begin{equation*}
     F(p_\mu):=  (1+\frac{ieaV_0}{2f(p_\mu)})^{-1}eaV_0(2\pi)^3
\end{equation*}
We can simplify these prefactors:
\begin{equation}\begin{aligned}
    \frac{F(p^{a,b}_\mu)\pi}{f(p^{a,b}_\mu)}&=\frac{-i eaV_0(2\pi)^3(1+\frac{ieaV_0}{2f(p^{a,b}_\mu)})^{-1}\pi}{f(p_\mu^{a,b})}\\
   &=\frac{-ieaV_0(2\pi)^3\pi}{f(p_\mu^{a,b})+\frac{ieaV_0}{2}}.
\end{aligned}\end{equation}
both of these functions take the form:
\begin{equation}\begin{aligned}
    \frac{1}{\sqrt{((p_\mu^x)^2-m^2)}+i\frac{eaV_0}{2} }
\end{aligned}\end{equation}
Since we have $p_\mu^a=(P+p)/2$ and $p_\mu^b=(P-p)/2$, the branch points for $p$ are given by the conditions on $p^a,p^b$ as
\begin{equation}\begin{aligned}
    \frac{E+p_0}{2}=&\pm \sqrt{(p_1^a)^2+(p_2^a)^2+m^2}\\\iff p_0=&\pm2\sqrt{(p_1^a)^2+(p_2^a)^2+m^2}-E\\
\end{aligned}\end{equation}
and
\begin{equation}\begin{aligned}
\frac{E-p_0}{2}=&\pm \sqrt{(p_1^a)^2+(p_2^a)^2+m^2}\\\iff p_0=&\pm2\sqrt{(p_1^a)^2+(p_2^a)^2+m^2}+E.\\
\end{aligned}\end{equation}
Therefore the branch cuts must emanate from $\pm(2\sqrt{(p_1^a)^2+(p_2^a)^2+m^2}-E)$ for $p_0$ in order to connect all the branch points. We also check whether there are poles of $p_0^x$ given by the condition $\sqrt{(p_0^x-m^2)}+i\frac{eaV_0}{2} =0$. This yields the condition that $p_0^x=\pm\sqrt{(p_1^a)^2+(p_2^a)^2+m^2-(eaV_0/2)^2}$ when $m^2-(eaV_0/2)^2>0$, and $p_0^x=\pm i \sqrt{(eaV_0/2)^2-(p_1^a)^2-(p_2^a)^2-m^2}$ when $m^2-(eaV_0/2)^2<0$. However, these are only solutions if we are on the second sheet of $f(p^x_0)$. Since only one of the square roots can be on the second sheet at once because $p^a_3=-p^b_3$, we neglect these poles. Accordingly, from now on let us fix sheets such that under on shell conditions, $f(p^a_\mu)=p^a_3$ and $f(p^b_\mu)=p^b_3$. The case where $(p_1^a)^2+(p_2^a)^2+m^2-(eaV_0/2)^2=0$ is also discarded to avoid the coincidence of the two poles.

It follows that the structure which we concern ourselves with of $\Gamma([P+p]/2)\Gamma([P-p]/2)$ as a function of $p_0$, consists of branch cuts emanating from $\pm(2\sqrt{(p_1^a)^2+(p_2^a)^2+m^2}-E)$ (negative NR energy) and the two usual poles given by the propagators at $-E\pm \sqrt{p_1^2+4m^2}$ and $+E\pm \sqrt{p_1^2+4m^2}$.

Overall, (11) has a very similar form to the 4D one-boson-exchange BS-equation treated originally by Wick and Cutkosky \cite{Wick1954}\cite{Cutkosky1954}. The only difference is that $\Gamma$ is no longer the bare propagator but one which contains a factor dependent on $\xi_\mu$, which should encapsulate tunneling effects. 

\section{Instantaneous Positive Energy Solution}
\subsection{Reduction to 1+1D}
Whether there is a closed form solution to (11) is still unknown. Nevertheless, we may extract a simpler 1+1D equation from (11) keeping only the energy dimension and tunneling dimension $x_3$ where the effects of the delta barrier is non trivial. This means replacing everywhere $\xi_\mu$ by $\xi_0$ and rewriting $f(\xi_0)=\sqrt{\xi_0^2-m^2}$. Let us suppress $P,U,u$ dependency for conciseness, and already in 1+1D, rewrite the first term as $G(p)$, and $\Gamma\left(\frac{P+p}{2}\right)\Gamma\left(\frac{P-p}{2}\right)$ as $C(p)$. Then, (11) is reduced to 
\begin{equation}
        L(p)=G(p)+g^2C(p)\int \frac{iL(v)}{(v-p')^2/4-\mu^2+i\epsilon}d^2v.
\end{equation}
To obtain a closed solution by Laplace transform, we may further make the instantaneous approximation
\begin{equation}\begin{aligned}
    \frac{i}{(v-p')^2/4-\mu^2+i\epsilon}\longrightarrow \frac{-i}{(v_1-p'_1)^2/4+\mu^2+i\epsilon}
\end{aligned}\end{equation}
the justification of which is explained in \cite{Petraki2015BoundStates}. Here we outline briefly. The kernel may be expanded as
\begin{align*}
    -\frac{4i}{A-(l_0)^2}=-\frac{4i}{A}\left(1+O\left(\frac{(l_0)^2}{A}\right)\right)
\end{align*}
where $A=(v_1-p'_1)^2+4\mu^2$ and $l_0=v_0-p'_0$.  In the NR limit we have $v_0\approx v_1^2/2m$ and $p_0'\approx p_1'^2/2m$. Then,
\begin{align*}
    l_0=O\left( \frac{v_1^2+p_1'^2}{2m}\right)=O\left(\frac{(v_1-p_1')(v_1+p_1')}{2m}\right).
\end{align*}
We also have $A=O((v_1-p_1')^2)$, therefore
\begin{align}
   \frac{(l_0)^2}{A}=O\left(\frac{(v_1+p_1')^2}{m^2}\right)
\end{align}
and we may keep only the first term if $L(v)$ dominates for ${(v_1+p_1')^2}/{m^2}$ small. Initially it may seem that this assumption must always be invalid because tunneling probability increases for larger momenta. However, we will see in section IV.A that for $g/eaV_0=O(10^{-3})$ and larger, interaction actually causes tunneling with low momenta to dominate. With a scaling parameter $v$ which we introduce in IV.B, (19) is just ${(l_0)^2}/{A}=O(v^2)$, hence the condition is that $L(v_1)$ should dominate for $v_1=O(mv)$.

With (18), we can replace the second term of (17) with a 1D integral. Defining $\left(\int L(v)dv_0\right):=L(v_1)$, $\int G(p)dp_0=G(p_1)$ and integrating both sides of (18) with $\int dp_0$ then gives 
\begin{equation}\begin{aligned}
   L(p_1)=G(p_1)&\\
   +g^2\int C(p) \int &\frac{-i}{(v_1-p'_1(p_0))^2/4+\mu^2-i\epsilon} L(v_1)dv_1dp_0,
\end{aligned}\end{equation}
where notably the one boson exchange is independent of the $p_0$ integral. Recall that $C(p)$ is a product of the $\Gamma_{\pm}$ factors, which contains pole structures described previously. We compute the $p_0$ integral by closing the contour in the lower half plane and picking up the particle poles of $-P_0+ \sqrt{p_1^2+4m^2}$ and $+P_0+\sqrt{p_1^2+4m^2}$. The residues of the free two particle propagator, multiplied with an arbitrary function $g(p_0)$ is then
\begin{widetext}
\begin{equation}\begin{aligned}
    &\text{Res } i\Delta (p^a) i\Delta(p^b)g(p_0)=\text{Res} \left(\frac{i}{(p^a)^2-m^2+i\epsilon}\frac{i}{(p^b)^2-m^2+i\epsilon}g(p_0)\right)\\
    =&2\pi i\frac{4 i}{2\sqrt{p_1^2+4m^2}}
    \frac{4i}{(2P_0-2\sqrt{p_1^2+4m^2})2P_0}g(-P_0+ \sqrt{p_1^2+4m^2})\\
    +&2\pi i
     \frac{4i}{(2P_0+2\sqrt{p_1^2+4m^2})2P_0}\frac{4 i}{-2\sqrt{p_1^2+4m^2}}g(P_0+ \sqrt{p_1^2+4m^2}).
\end{aligned}\end{equation}
\end{widetext}
This is the standard (relative) energy residue for a free two particle propagator, where the restriction of $p_0$ to $-P_0+ \sqrt{p_1^2+4m^2}$ and $P_0+ \sqrt{p_1^2+4m^2}$ correspond to mutually incompatible on shell conditions. Namely,
\begin{equation}\begin{aligned}
    p_0+P_0=+ \sqrt{p_1^2+4m^2} \implies &(p_0+P_0)^2-p_1^2=4m^2\\
    &\iff (p^a)^2=m^2
\end{aligned}\end{equation}
\begin{equation}\begin{aligned}
    p_0-P_0=+ \sqrt{p_1^2+4m^2} \implies &(p_0-P_0)^2-p_1^2=4m^2\\
    &\iff (p^b)^2=m^2
\end{aligned}\end{equation}
and
\begin{equation}\begin{aligned}
      p_0+P_0=+ \sqrt{p_1^2+4m^2}\; \wedge\; p_0-P_0=+ \sqrt{p_1^2+4m^2} \\
      \implies -P_0=P_0\implies P_0=0
\end{aligned}\end{equation}
Since we need both $p^a$ and $p^b$ to be on shell at the same time in order to have $p'_1=f(p_0^a)-f(p_0^b)=p^a_1-p_1^b=p_1$, which is necessary if we want to solve (20) in closed form, we must neglect one of the terms. Neglecting the negative energy term which is the third line of (21), is a canonical approximation for the instantaneous (Salpeter) reduction with a positive-energy / no-pair projection (and closely related quasipotential reductions such as Blankenbecler–Sugar, Thompson, and Gross) \cite{BetheSalpeter1951,BlankenbeclerSugar1966,Thompson1970,Gross1969}. Through computation included in appendix C, we may show that this approximation is also controlled by the assumption that $L(v)$ dominates for small momenta, hence interaction encourages low momenta tunneling. The imposed on shell condition for $p_b$ and (23) then implies $p_0=P_0-\sqrt{p_1^2+4m^2}\implies p^b_0=\sqrt{p_1^2+4m^2} $, as well as $P'=P$, which gives us the reduced but solvable form of 20:
\begin{widetext}
    \begin{equation}
        L(p_1)=G(p_1)-ig^2C(p_1)\int\frac{1}{(v_1-p_1)^2/4+\mu^2-i\epsilon}L(v_1)dv_1
    \end{equation}
where, with $E_{p_1}=\sqrt{p_1^2+4m^2}$,
\begin{equation}
    C(p_1)=2\pi i\frac{2 i}{E_{p_1}}
    \frac{i}{(P_0-E_{p_1})P_0}\left(1+\frac{-ieaV_0(2\pi)\pi}{ p^a_1+\frac{ieaV_0}{2}}\right)\left(1-\frac{-ieaV_0(2\pi)\pi}{p^b_1+\frac{ieaV_0}{2}}\right).
\end{equation}
\end{widetext}
\subsection{Closed Form Solution in Laplace Space}
A non perturbative, closed form approximate solution exists for (25) via the generalized Wiener Hopf technique. Because this procedure is algebraically extensive and mathematically self-contained, its detailed implementation warrants a separate paper. Here we will outline how (25) may be solved.\\

An analytic solution to (25) rests on properties of (26), which we have already written in factorized form as $M(p_1)R(p_1)$ consisting of an even function $M$ and rational factors $R$. We denote their Fourier transform as ${m}$ and ${r}$ respectively. Notably, 
\begin{equation}
r(x)=\delta(x)+\alpha\!\left[i\,e^{\gamma x}\Theta(-x)-i\,e^{-\gamma x}\Theta(x)\right]
+\frac{\alpha^{2}}{2\gamma}\,e^{-\gamma|x|}.
\end{equation}
Where $\alpha=-ieaV_0(2\pi)^3\pi$, and $\gamma=\frac{eaV_0}{2}$ are constants. Let us use lower case characters to denote the Fourier conjugates. After inverse Fourier transform, (25) takes the form
\begin{equation}
    l(x)=g(x)-ig^2(m\ast r\ast s)(x)
\end{equation}
Where we have denoted convolution of the real line as $\ast$, and $s(x)=kl(x)$, with $k$ the one boson kernel. Then we define for arbitrary Fourier transformed function $f$, the one sided functions $f_\pm(x)=\Theta(\pm x)f(x)$, and the Laplace transform $\tilde{F}_\pm (s):=\int_0^\infty e^{-sx}f_{\pm}(\pm x)dx=\mathcal{L}_\pm[f](s)$. Taking the Laplace transform on both sides we obtain
\begin{equation}
\tilde{L}_\pm(s)=\tilde{G}_\pm(s)-ig^2\mathcal{L}_\pm[(m\ast r\ast s)](s).
\end{equation}
The decoupling of $\mathcal{L}_\pm[(m\ast r\ast s)](s)$ is non-trivial because $\ast$ is a convolution over the full line. We therefore need to write $m$ in terms of $m_\pm$ and same for $(r\ast s)$. Let us also denote by $\star$ convolution over the half line (which half is made obvious by context). Since $a*b=a_+\star b_+ + a_+\ast b_- + a_-\ast b_++a_-\star b_-$, we have that
\begin{widetext}
\begin{equation}\begin{aligned}
    &\mathcal{L}_\pm[(m\ast r\ast s)](s)\\
    =&\mathcal{L}_\pm[m_+\star (r\ast s)_+ + m_+\ast (r\ast s)_- + m_-\ast (r\ast s)_++m_-\star (r\ast s)_-]\\
    =&\mathcal{L}_\pm[m_\pm\star (r\ast s)_\pm + m_+\ast (r\ast s)_- + m_-\ast (r\ast s)_+]\\
    =&\mathcal{L}_\pm[m_\pm\star (r\ast s)_\pm ]+\mathcal{L}_\pm[ m_+\ast (r\ast s)_- + m_-\ast (r\ast s)_+]\\
    =&\tilde{M}_\pm (s) \mathcal{L}_\pm[(r\ast s)] (s)+\mathcal{L}_\pm[ m_+\ast (r\ast s)_- + m_-\ast (r\ast s)_+].
\end{aligned}\end{equation}
\end{widetext}
Because $r$ contains only Volterra integrals, standard Laplace identities apply allowing us to decouple $r$ and $s$. In particular we get the form 
\begin{equation}\begin{aligned}
   \mathcal{L}_\pm[r\ast s]=A(s)S_\pm(s)+ C(s)S_\pm(\gamma)+D(s)S_\mp(\gamma),
\end{aligned}\end{equation}
where importantly $S_\mp(s)$ does not appear because the opposite–half pieces of $r$ are one-sided Volterra kernels whose unilateral Laplace transforms yield only boundary evaluations at the simple poles $s=\pm\gamma$ (i.e.,  $S_\mp(\pm\gamma)$ , not an $s$-dependent factor $S_\mp(s)$). 

$\mathcal{L}[m_\pm\ast (r\ast s)_\mp]$ does not decouple without further approximation. This requires us to take $m_\pm\approx m_\pm(0)\delta(x)$. Since $m$ does decouple for the other terms, we keep the full expression there. Then, (30) becomes
\begin{widetext}\begin{equation}\begin{aligned}
        \mathcal{L}_\pm[(m\ast r\ast s)](s)& =(\tilde{M}_\pm(s)+m_\mp(0))\mathcal{L}_\pm[(r\ast s)]\\
        &=(\tilde{M}(s)+m(0))\times(A(s)S_\pm(s)
   +C(s)S_\pm(\gamma)+D(s)S_\mp(\gamma)) .\\
\end{aligned}\end{equation}\end{widetext}
Where the last equation follows from $m$ being even. Let $\mu'$ denote $\sqrt{\mu^2+i\epsilon}$. Since we also have by definition
\begin{equation}
    \tilde{S}_\pm(s)=\frac{1}{\mu'}\tilde{L}_\pm(s+2\mu'),
\end{equation}
our overall system of equation, denoting $\mathcal{M}(s):=\tilde{M}(s)+m(0)$, becomes 
\begin{widetext}\begin{equation}\begin{aligned}
        \tilde{L}_+(s)=\tilde{G}_+(s)-ig^2\mathcal{M}(s)\frac{1}{\mu'}\left(A(s)\tilde{L}_+(s+2\mu')+C(s)\tilde{L}_+(\gamma+2\mu')+D(s)\tilde{L}_-(\gamma+2\mu')\right)\\
        \tilde{L}_-(s)=\tilde{G}_-(s)-ig^2\mathcal{M}(s)\frac{1}{\mu'}\left(A(s)\tilde{L}_-(s+2\mu')+C(s)\tilde{L}_-(\gamma+2\mu')+D(s)\tilde{L}_+(\gamma+2\mu')\right)
    \end{aligned}\end{equation}\end{widetext}
Let us define the shift operator 
$\mathsf S_{2\mu'}L_\pm(s):=L_\pm(s+2\mu')$, and the matrices

\begin{align}
\mathsf L(s)&:=
\begin{pmatrix}
    \tilde{L}_+(s)\\
    \tilde{L}_-(s)
\end{pmatrix},\\[4pt]
\mathsf K (s)&:=
\begin{pmatrix}
    \tilde{L}_+(\gamma+2\mu')\\
    \tilde{L}_-(\gamma+2\mu')
\end{pmatrix},\\[4pt]
\mathsf G (s)&:=
\begin{pmatrix}
    \tilde{G}_+(s)\\
    \tilde{G}_-(s)
\end{pmatrix}\\[4pt]
\mathsf A(s) &:=
\frac{i g^2 \mathcal{M}(s)}{\mu'}\,A(s)\,
\begin{pmatrix}
\mathsf S_{2\mu'} &0\\[2pt]
0 & \mathsf S_{2\mu'}
\end{pmatrix},
\\[4pt]
\mathsf B(s) &:=
\frac{i g^2 \mathcal{M}(s)}{\mu'}\,
\begin{pmatrix}
C(s) & D(s)\\[2pt]
D(s) & C(s)
\end{pmatrix}.
\end{align}
Then, (34) is just
\begin{equation}
\begin{aligned}
    \mathsf L=\mathsf G+\mathsf A\mathsf L+\mathsf B\mathsf K\\
    \iff(1-\mathsf A)\mathsf L=\mathsf G +\mathsf B\mathsf K\\
    \mathsf L=(1-\mathsf A)^{-1}(\mathsf G+\mathsf B \mathsf K)
\end{aligned}
\end{equation}
It remains to compute $(1-\mathsf{A})^{-1}$, if it exists. Since $(1-\mathsf{A})$ is diagonal, it suffices to compute an inverse for the functional operator $1-\mathcal{C}S_{2\mu'}$ for some constant $\mathcal{C}(s)$. No inverse exists for this operator except if we approximate $\mu'$ to be small and take $S_{2\mu'}L(s)= \tilde{L}(s)+\mu'\partial_s\tilde{L}(s)+O(\mu'^2)$. Physically, this assumes a small meson mass. In fact, if the boson mass is set to be infinitesimally small, we will recover the coulomb interaction in the NR reduction. This raises an interesting point: Whilst the coulomb interaction is long-range and therefore thwarts short-range scattering formalism in quantum mechanics, approaching the coulomb interaction with an asymptotically small meson massed Yukawa interaction is essential to a solution in our field theoretic formulation. Also, as the ladder approximation is justified for $|g/eaV_0|\ll1$ or for large $\mu$ or in the NR regime, to look for a non-perturbative approximation with a small $\mu$ means we must solve in the NR regime. With this approximation, the operator gives
\begin{equation}
\begin{aligned}
  \Phi(s):&=(1-\mathcal{C}S_{2\mu'})L(s)\\
  &= (1-\mathcal{C})\tilde{L}(s)-2\mathcal{C}\mu'\partial_s\tilde{L}(s)+O(\mu'^2),
  \end{aligned}
\end{equation}
\vspace{-0.1cm}
which implies
\begin{equation}
    \begin{aligned}
          (1+ \mathcal{K}S_{2\mu'})\Phi&=((1+\mathcal{K})\Phi+2\mathcal{K}\mu'\partial_s\Phi)+O(\mu'^2)\\
  &=(1+\mathcal{K})\left[(1-\mathcal{C})\tilde{L}(s)-2\mathcal{C}\mu'\partial_s\tilde{L}(s)\right]\\
  &\;+2\mathcal{K}\mu'(1-\mathcal{C})\partial_s\tilde{L})+O(\mu'^2)\\
    \end{aligned}
\end{equation}
The derivative terms cancel out if we select $\mathcal{K}(1-\mathcal{C})=(1+\mathcal{K})\mathcal{C}\iff\mathcal{K}=\mathcal{C}/(1-2\mathcal{C})$. Normalizing then gives us, to leading order in $\mu'$:
\begin{equation}
    (1-\mathsf{A})^{-1}=\frac{1}{(1+\mathcal{K})(1-\mathcal{C})}\left[1+\mathcal{K}(s)\begin{pmatrix}
\mathsf S_{2\mu'} &0\\[2pt]
0 & \mathsf S_{2\mu'}
\end{pmatrix}\right],
\end{equation}
where 
\begin{equation}
    \mathcal{C}(s)=\frac{i g^2 \mathcal{M}(s)}{\mu'}\,A(s).
\end{equation}

Given the boundary conditions $\tilde{L}_{\pm}(\gamma+2\mu')$, everything on the RHS of (40) is now known. We have arrived at a closed form solution in Laplacian space. To obtain $l$, we perform an inverse Laplace transform. Then to obtain the original $L$ in momentum space, we Fourier transform $l$. \\

It must be pointed out that the method of splitting the functions into positive and negative support parts in $x$ space corresponds to the symmetry of the system. Although $x$ is the relative distance in the tunneling direction, since we fix $P_0$, the center of mass $X$ is completely de-localized. Therefore, every $P_0,x$ state in fact encodes a superposition of all $X,p$ states, which includes the case where $X=0$. $x=0$ Then necessarily involves the physical discontinuity we have introduced at $x^a,x^b=0$. Hence, splitting at $x=0$ was a natural choice well reflected in the structure of $r$.

\section{Physical Consistency and Relevancy}
Whilst (5) was shown to be consistent with relativistic quantum mechanics \cite{Zielinski2024b}, whether the formalism of (8) is relevant to conventional two particle tunneling problems is uncertain. In this section we derive the tunneling amplitude by perturbing with respect to $|\frac{g}{eaV_0}|$ to examine physical properties, and bridge our field theoretic formalism with conventional tunneling problems by reducing (8) in the NR limit to a Lippmann-Schwinger equation corresponding to a Hamiltonian of class (1). These preliminary studies will justify further pursuits of non-perturbative solutions for (8).
\hspace{4cm}
\subsection{Born Approximation for Tunneling Amplitude}
For simplicity, we remain in the 1+1D case as outlined previously. Let us compute the second term of the RHS of (7), which, without the $-g^2$, reads
\begin{widetext}
\begin{equation}\begin{aligned}
    B_1(p^a,p^b;u^a,u^b)=\int &D_\psi(p^a,k^a)D_\psi(p^b,k^b)D_{\phi}(q) D_\psi(v^a,u^a)D_\psi(v^b,u^b)\\
    &\times \delta^2(v^a-k^a-q)\delta^2(v^b-k^b+q)d^2k^ad^2k^bd^2v^ad^2v^bd^2q.
\end{aligned}\end{equation}
\end{widetext}
Using the same notation as defined in (10), integrating over $q$ gives momentum conservation $\delta^2(V-K)$. Then,  integrating over $V$ collapses this delta to give us
{\setlength{\abovedisplayskip}{0pt}%
 \setlength{\abovedisplayshortskip}{0pt}%
\begin{equation}\begin{aligned}
     &B_1(P,p;U,u)=\int D_\psi(p^a,k^a)D_\psi(p^b,k^b)D_{\phi}\left(\frac{k-v}{2}
     \right) \\
     &\times D_\psi((K+v)/2,u^a)D_\psi((K-v)/2,u^b) d^2k^a\,d^2k^b\,d^2v.
\end{aligned}\end{equation}}
We then use energy conservation in $D_\psi$ on the out legs to integrate out $k^a_0,k^b_0$ which become $p^a_0,p^b_0$. Then we are left with pole structures which also integrate out (see appendix B, here we take the positive pole for $p^b$ since we are in an arbitrary frame). If we further grant on shell conditions  for $p^a,p^b$ we have $f(p^a_0)=p^a_1$ and  $f(p^b_0)=p^b_1$. However since we have not amputated, we will abuse notation for conciseness and write $\sqrt{(p_0^a)^2-m^2}=p_1'^a$ whilst suppressing the "$'$", with the understanding that we don't actually go on shell yet. After we amputate and go on shell, these equals will be realized. 

With this abuse of notation, the result is
\begin{equation}
    \begin{aligned}
&B_1(P,p;U,u)=\int\Gamma(p^a)\Gamma(p^b)D_\phi\left(\frac{p-v}{2}\right)\\
&\times D_\psi((P+v)/2,u^a)D_\psi((P-v)/2,u^b) d^2v.
    \end{aligned}
\end{equation}
We may then use the same method to integrate out $v$. Doing the $v_0$ integral first gives us a simultaneous condition $(P_0+v_0)/2=u^a_0$ and $(P_0-v_0)/2=u^b_0$. This gives a factor of $\delta(P_0-U_0)$ which we factor out, unsurprising because our model retains translational symmetry in time coordinate. Then, $D_\psi D_\psi$ could be expanded into four terms. The first (trivial) term
\begin{equation}\begin{aligned}
\frac{i}{(P+v)^2/4-m^2+i\epsilon}\frac{i}{(P-v)^2/4-m^2+i\epsilon}\\
\times\delta^1((P_1+v_1)/2-u_1^a) \delta^1((P_1-v_1)/2-u_1^b).
\end{aligned}\end{equation}
Gives the full simultaneous condition $(P+v)/2=u^a$ and $(P-v)/2=u^b$. Without thinking we know that it becomes
\begin{equation}
i\Delta(u^a)i\Delta(u^b)\times\delta^1(P_1-U_1) .
\end{equation}
The second term is
\begin{equation}\begin{aligned}
     &i\Delta((P+v)/2)\delta^1((P_1+v_1)/2-u_1^a) \\
     \times &i\Delta((P-v)/2)F((P-v)/2)i\Delta(u^b).
\end{aligned}
\end{equation}
The delta function forces this to become
\begin{equation}\begin{aligned}
     i\Delta(u^a)i\Delta(P-u^a)F(P-u^a)i\Delta(u^b).
\end{aligned}
\end{equation}
The third term collapses symmetrically to become
\begin{equation}\begin{aligned}
     i\Delta(u^b)i\Delta(P-u^b)F(P-u^b)i\Delta(u^a).
\end{aligned}
\end{equation}

Contour integration is required for the fourth term. the pole structure for $v_1$ reads
\begin{equation}\begin{aligned}
    B_1=\int&\frac{4i}{(2u^a_0)^2-(P_1+v_1)^2-4m^2+i\epsilon}\\
    \times&\frac{4i}{(2u_0^b)^2-(P_1-v_1)^2-4m^2+i\epsilon}(...)(v_1)dv_1
\end{aligned}
\end{equation}
with corresponding poles at $v_1=P_1\mp2\sqrt{(P_0-u_0)^2/4-m^2}\mp i\epsilon$, or $v_1=-P_1\pm2\sqrt{(P_0+u_0)^2/4-m^2}\mp i\epsilon$ (we suppress the rest of the $v_1$ dependent integrand as $(...)(v_1)$). However we have just established that $P_0=U_0$, so $P_0+u_0=U_0+u_0=2u^a_0$, and $P_0-u_0=U_0-u_0=2u^b_0$. If we take $u^a,u^b$ to be on shell we are further granted $\sqrt{(u^a_0)^2-m^2}=u_1^a$ and $\sqrt{(u^b_0)^2-m^2}=u_1^b$. We do the same abuse of notation as before. If we then close the contour in the lower half plane for $v_1$, hence picking $P_1+v_1=2u_1^a,P_1-v_1=2u_1^b$, we obtain
\begin{equation}\begin{aligned}
B_1=\int& \left(\frac{2\pi}{u_1^a}
    \frac{i}{(u_1^b)^2-(P_1-u_1^a)^2+i\epsilon}\right)(...)(-P_1+2u^a_1)\\
    + &\left(\frac{i}{(u_1^a)^2-(P_1-u_1^b)^2+i\epsilon}\frac{2\pi}{u_1^b}\right)(...)(P_1-2u^b_1)du_1
\end{aligned}
\end{equation}
\vspace{-1pt}
Notice that for $v_1=-P_1+2u^a_1$, $p-v$ just becomes $2(p^a-u^a)$. For $v_1=-P_1+2u^a_1$, $p-v$ becomes $2(u^b-p^b)$. Therefore, $B_1$ decomposes into 5 terms
\begin{widetext}   
\begin{equation}\begin{aligned}B_1=\Gamma(p^a)\Gamma(p^b)\delta(u^a_0+u^b_0-p^a_0+p^b_0)\bigg( &D_\phi(p^a+u^b-p^b-u^a) i\Delta(u^a)i\Delta(u^b)\delta(u^a_1+u^b_1-p^a_1-p^b_1)\\
    + &D_\phi(p^a-u^a)i\Delta(u^a)i\Delta(P-u^a)F(P-u^a)i\Delta(u^b)\\
    +&D_\phi(p^b-u^b)     i\Delta(u^b)i\Delta(P-u^b)F(P-u^b)i\Delta(u^a)\\
    +&D_\phi(p^a-u^a)\left(\frac{2\pi}{u_1^a}
    \frac{i}{(u_1^b)^2-(P_1-u_1^a)^2+i\epsilon}\right)F(u^a_0)F(u^b_0)i\Delta(u^a)i\Delta(u^b)\\
    +&D_\phi(p^b-u^b)\left(\frac{i}{(u_1^a)^2-(P_1-u_1^b)^2+i\epsilon}\frac{2\pi}{u_1^b}\right)F(u^a_0)F(u^b_0)i\Delta(u^a)i\Delta(u^b)\bigg)\\
\end{aligned}
\end{equation}
\end{widetext}
Define $\Lambda$ as
\begin{equation}
    \Lambda(p_1^a)=1+\frac{-ieaV_0(2\pi)}{ p_1^a+ieaV_0},
\end{equation}
Hence $\Gamma(p^a)\Gamma(p^b)=i\Delta(p^a)i\Delta(p^b)\Lambda(p^a_1)\Lambda(p^b_1)$.
Since the first term belongs to the $\delta^2$ (non tunneling) sector, we call it $B_0$ and denote everything else in the brackets separately as $\Omega(p^a,p^b;u^a,u^b)i\Delta(u^a)i\Delta(u^b)$.  We therefore take $B_1$ \textit{less the first term}, amputate it, and take all coordinates to be actually on shell. This gives us the tunneling $T$ matrix element $\mathcal{T}$, as
\begin{equation}
    i\mathcal{T}(p^a,p^b;u^a,u^b)=-g^2\Lambda(p^a_1)\Lambda(p^b_1)\Omega(p^a,p^b;u^a,u^b).
\end{equation}
The S-matrix is then given by
\begin{equation}
    \begin{aligned}
    \braket{p^ap^b|S|u^au^b}&=\delta^2(p^a-u^a)\delta^2(p^b-u^b)(\text{free term})\\
    +&\delta^2(U-P)B_0\\
    +&\delta(p^a_0-u^a_0)\delta(p^b_0-u^b_0)F(p^a)F(p^b)\\
    -&\delta(U_0-P_0)  i\mathcal{T}(p^a,p^b;u^a,u^b),
\end{aligned}
\end{equation}
where $\ket{p^ap^b}$ and $\ket{u^au^b}$ are the momentum Fock states, and $p^a=[E_{p^a_1},p^a_1]$. The $F$ here, corresponding to the first term of the RHS of (7), is same as the $F$ used previously, being the S-Matrix element derived in \cite{Zielinski2024b}. The on-shell condition now allows it to take its original form:
\begin{equation}
    F(p^a)=  -ieaV_0(2\pi)\left(1+\frac{ieaV_0}{2p_1^a}\right)^{-1}.
\end{equation}
Only the third and fourth term of (58) include $u(x)$ insertions, and since they are in different delta sectors, one must smear against wave packets in order to compare their contribution quantitatively. This is implemented with the mollification $\delta(x)\rightarrow \frac{1}{\pi}\frac{\epsilon}{x^2+\epsilon^2}$, and gaussian wave packets of the form
\[
\phi_{p_1'}(p_1) = N(p_1') \exp\!\left(-\frac{(p_1 - p_1')^2}{4\sigma^2}\right),
\]
where $N(p_1')$ is the normalization with respect to the invariant measure. Define $t(p_1'^a,p_1'^b)$ as the smeared result with respect to incoming and outgoing wave packets $\phi_{p_1'^a}\phi_{p_2'^a}$ both centered at the same two momenta $p_1'^a, p_1'^b$. 
\begin{figure*}[t]
\centering
\subfloat[$g/eaV_0=1$]{\includegraphics[width=0.5\textwidth]{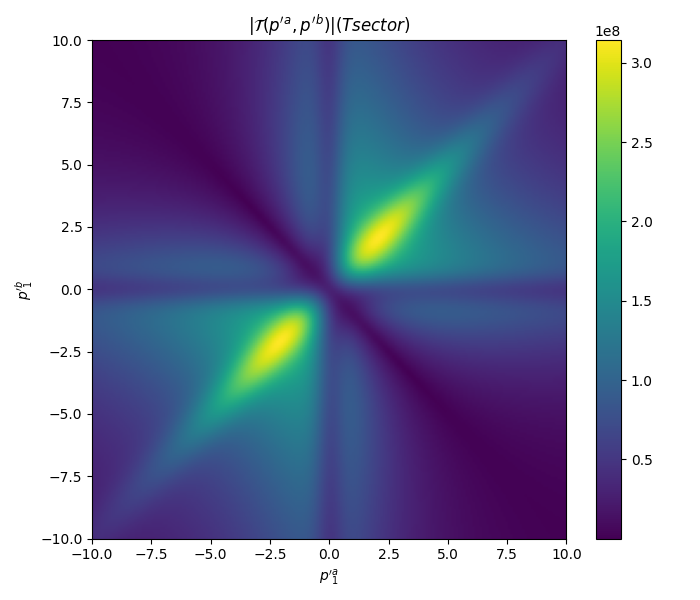}}
\subfloat[$g/eaV_0=10^{-3}$]{\includegraphics[width=0.5\textwidth]{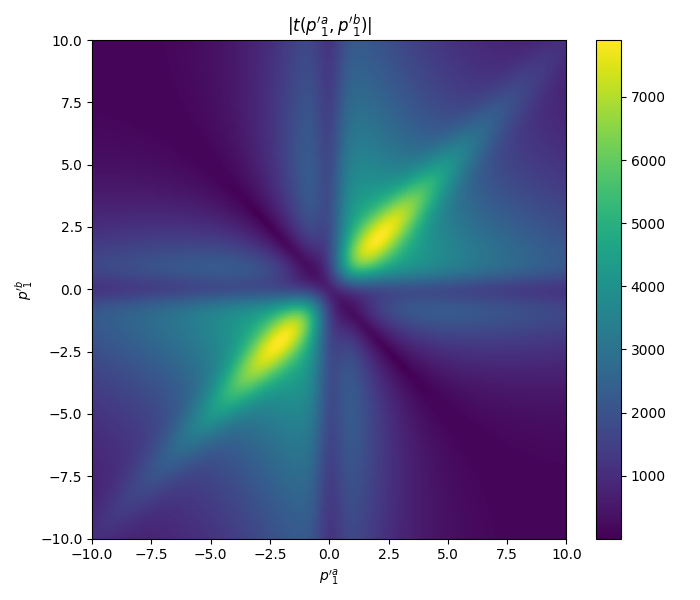}}\\
\subfloat[$g/eaV_0=10^{-4}$]{\includegraphics[width=0.5\textwidth]{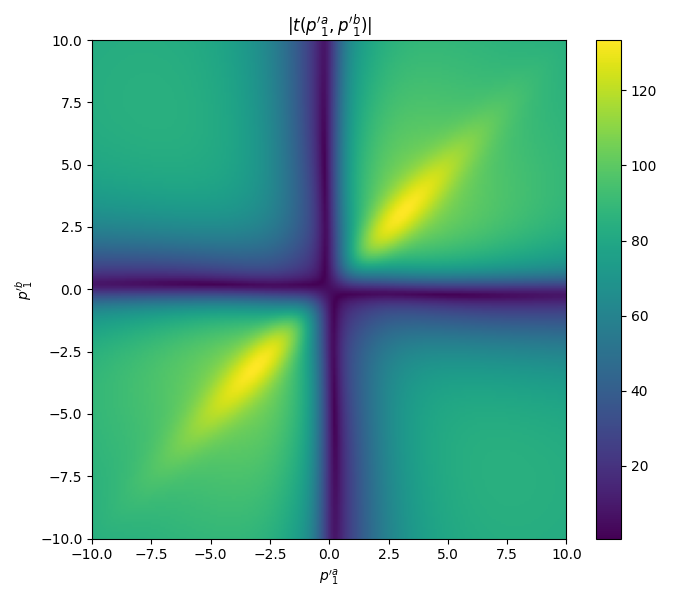}}
\subfloat[$g/eaV_0=10^{-5}$]{\includegraphics[width=0.5\textwidth]{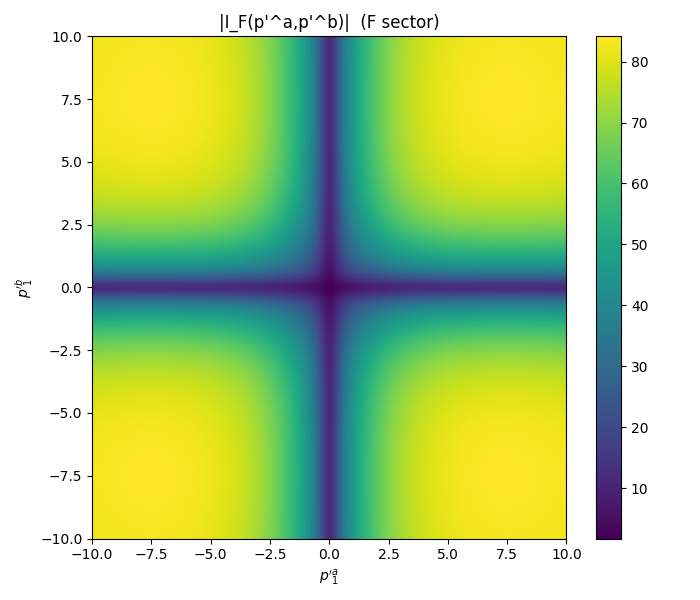}}
\caption{Heat map of smeared $S$-matrix $|t|$ as a function of $p_1^{\prime a},p_1^{\prime b}$ with mollified delta functions. Parameters: $m = 25,\,\mu = 1,\ eaV_0 = 5,\,\epsilon = 0.1$. The large values on the color axis arise from pole-like structures of $\mathcal{T}$, notably the $2\pi/u_1^{a,b}$ terms in the third and fourth term of (55).}
\label{fig:smearedS}
\end{figure*}

The heat map of $|t|$ is shown in FIG. 1. for different values of $g/eaV_0$. Importantly, we may verify that for larger values of $g/eaV_0$, the dominant contribution comes from smaller momenta, hence justifying our approximation for the non-perturbative solution, the relevance of which is precisely in the case $g/eaV_0\gg1$. More specifically, the $T$-matrix, shown in FIG. 1(a), begins to dominate at $g/eaV_0=O(10^{-3})$, below which the $FF$ term takes effect and we get back the non-interacting tunneling case. Furthermore, the $T$-matrix has very large values due to the pole like structures contained in the third and fourth term of (55).

There is a ridge along the diagonal $(p_1'^a \simeq p_1'^b)$ and a suppression band along the anti-diagonal $(p_1'^a \simeq -\,p_1'^b)$, most pronounced at low momenta, with an asymmetry between same-sign and opposite-sign quadrants. These features originate from the exchange-type contributions (the third and fourth terms of Eq.~(55)). On the anti-diagonal $p_1'^a \simeq -\,p_1'^b$ so $P_1 \simeq 0$, and the corresponding denominators satisfy
\begin{equation}\begin{aligned}
(p_1^b)^2-(P_1-p_1^a)^2 
&\simeq (p_1^b)^2-(p_1^a)^2\\
= -\big[(p_1^a)^2-(p_1^b)^2\big]
&\simeq -\big[(p_1^a)^2-(P_1-p_1^b)^2\big].  
\end{aligned}\end{equation}
Since the two terms are related by $a \leftrightarrow b$, their contributions cancel to a good approximation under the (approximately) exchange-symmetric smearing, producing suppression near $P_1 \simeq 0$. By contrast, on the diagonal $p_1'^a \simeq p_1'^b$ so $P_1 \simeq 2p$ and
\begin{equation}
(p_1^b)^2-(P_1-p_1^a)^2 \simeq p^2-p^2 \simeq 0,
\
(p_1^a)^2-(P_1-p_1^b)^2 \simeq 0,
\end{equation}
so both denominators approach their pole surfaces, yielding an enhanced ridge.

Therefore in the neighborhood of $(0,0)$, we see that tunneling is generally promoted by interaction except for completely opposite momenta, where the effect of interaction is completely suppressed. This is indicative of destructive interference between the forward- and backward-propagating two-body virtual states, which cancels the correlated contribution when the total momentum vanishes. Interestingly, this raises questions as to whether effects of interaction on tunneling could be observed for a non-perturbative solution obtained via the method outlined in III.B, where we imposed that the total momentum is $0$. Nevertheless, since we were not required to choose $u_1'^{a,b}=p_1'^{a,b}$, this suppression may not manifest if we only place the out-coordinates on the anti-diagonal.

The asymmetry in mechanism between same and differing sign momenta could well be a clue to the frequency splitting found in the numerical study of systems like (1) \cite{Brugger2025}, where a completely symmetric system of two identical bosons produced two distinct tunneling frequencies.

\subsection{Reduction in NR Limit}
To facilitate the transform into a Lippmann-Schwinger equation, we perform two steps. First, we integrate by $\int dp_0$ on both sides. Secondly, we take the instantaneous approximation for the one boson kernel:
\begin{equation}\begin{aligned}
    \frac{i}{(v-p')^2/4-\mu^2+i\epsilon}\longrightarrow \frac{-i}{(\mathbf{v}-\mathbf{p'})^2/4+\mu^2}
\end{aligned}\end{equation}
the justification of which in the NR limit we have already discussed when deriving (19). Since this procedure is similar to the one in section III.A., the specifics are moved to appendix. Where the procedure begins to differ is when we take the NR limit  $\frac{|\mathbf{p}|}{m}\ll1$.\\

Let us define the NR energy as $\mathcal{E}=E-2m$. then the denominator of (21), now in full 4D, corresponds to
\vspace{-2pt}
\begin{equation}\begin{aligned}
    \sqrt{\mathbf{p}^2+4m^2}&=2m+\frac{\mathbf{p}^2}{4m}+O(\mathbf{p}^4/m^3)\\
    2E-2\sqrt{\mathbf{p}^2+4m^2}&=2\mathcal{E}-\frac{\mathbf{p}^2}{2m}+O(\mathbf{p}^4/m^3)\\
        2E+2\sqrt{\mathbf{p}^2+4m^2}&=8m+2\mathcal{E}+\frac{\mathbf{p}^2}{2m}+O(\mathbf{p}^4/m^3).
\end{aligned}\end{equation}
The leading term of the denominator of the second term is therefore $128m^3$, which means we can drop it for $|\mathbf{p}|$ sufficiently small. To complete the passage to the NR regime, we must also restrict that $\mathcal{E}\ll m$. To moderate the double limit we define a parameter $v$ where $|\mathbf{p}|\propto mv$, $\mathcal{E}\propto mv^2$, and $mv\rightarrow0$. Then the simplified 1st term is
\begin{widetext}
\begin{equation}\begin{aligned}
    \frac{4\pi i}{E\sqrt{\mathbf{p}^2+4m^2}(\sqrt{\mathbf{p}^2+4m^2}-E)}g(-E+ \sqrt{\mathbf{p}^2+4m^2})\\
    =\frac{4\pi i}{[4m^2+O(m^2v^2)][-\mathcal{E}+\frac{\mathbf{p}^2}{4m}+O(mv^4)]}g(-E+ \sqrt{\mathbf{p}^2+4m^2})\\
    =\frac{-4\pi i}{4m^2[\mathcal{E}-\frac{\mathbf{p}^2}{4m}]}g(-E+ \sqrt{\mathbf{p}^2+4m^2})+O(m^3v^4)
\end{aligned}\end{equation}
\end{widetext}
And we have obtained the Schrodinger resolvent, coupled with the on shell condition for the rest of the function sending $P_0+p_0\rightarrow  \sqrt{\mathbf{p}^2+4m^2}$. The latter has the further implications that 
\begin{equation}\begin{aligned}
    (P_0+p_0)^2=\mathbf{p}^2+4m^2\\
    \iff(P_0+p_0)^2-p_1^2-p_2^2-m^4/4=p_3^2\\
    \iff\pm\sqrt{[(P_0+p_0)^2-p_1^2-p_2^2]/4-m^2}=p_3/2\\
    \iff \pm f(p^a_\mu)=p^a_3.
\end{aligned}\end{equation}
However since we have by construction $p_3^a+p_3^b=0$, we are further granted $p_3^b=-p_3^a=\mp f(p^a_\mu)$. In order to recover a true Yukawa potential, we further need $\pm f(p_\mu^b)=p_3^b$. We may recover this by comparing $f(p_0^a)$ and $f(p_0^b)$, and observing that $p^b$ is only mildly off shell in the NR limit
\begin{widetext}
\begin{equation}\begin{aligned}
    f(p_\mu^b)&=\sqrt{f(p_\mu^a)^2-Ep_0}=\sqrt{(-p_3^b)^2\left(1-\frac{Ep_0}{(p_3^b)^2}\right)}=\sqrt{(p_3^b)^2}\sqrt{1-\frac{Ep_0}{(p_3^b)^2}}\\
    &=p_3^b(1-\frac{Ep_0}{2(p_3^b)^2}+O(mv^2))\\
   & =p_3^b+\frac{Ep_0}{2p_3^b}+O(m^2v^3)=-p_3^b+\frac{Ep_0}{E-p_3}+O(m^2v^3)\\
   &=p_3^b+\frac{E(-\mathcal{E}+p_3^2/4m)}{E-p_3}+O(mv^4)+O(m^2v^3)\\
   &=p_3^b+\frac{E(-\mathcal{E}+p_3^2/4m)}{E-p_3}+O(m^2v^3)\\
   &=p_3^b+O(mv^2)+O(m^2v^3)\\
   &=p_3^b+O(mv^2).
\end{aligned}\end{equation}

\end{widetext}
Therefore inside $g$, $p^b$ is basically also on shell. Furthermore, by fixing $p_1^a>0$, the COM condition implies
$p_3^a=-p^b_3$, hence $p_3^b<0$. Therefore for $p^b$ on shell, $p_3^b$ is nothing but $-f(p_0^b)$ (with correction at $O(mv^2)$). The one boson exchange kernel can then be approximated to
\begin{widetext}
\begin{equation}\begin{aligned}
    \frac{-i}{(v_1-p_1)^2/4+(v_2-p_2)^2/4+(v_3-[p_3^a+f(p_\mu^b)])^2/4+\mu^2}\\=\frac{-i}{(v_1-p_1)^2/4+(v_2-p_2)^2/4+(v_3-[p_3^a-p_3^b])^2/4+\mu^2}+O(mv^2)\\=\frac{-i}{(\mathbf{v}-\mathbf{p})^2/4-\mu^2}
+O(mv^2)\end{aligned}\end{equation}
\end{widetext}
However we know that ${-4\pi i}/{4m^2[\mathcal{E}-\frac{\mathbf{p}^2}{4m}]}\times g(-E+ \sqrt{p_1^2+4m^2})$ is already order $O(m^3v^2)$, therefore, with $h$ denoting the remaining prefactors,
\begin{widetext}
\begin{equation}\begin{aligned}
    \left(\frac{-4\pi i}{4m^2[\mathcal{E}-\frac{\mathbf{p}^2}{4m}]}h(p^a,p^b\rightarrow\text{on shell})+O(m^3v^4)\right)\left(\frac{-i}{(\mathbf{v}-\mathbf{p})^2/4-\mu^2}
+O(mv^2)\right)\\
= \frac{-4\pi i}{4m^2[\mathcal{E}-\frac{\mathbf{p}^2}{4m}]}h(p^a,p^b\rightarrow\text{on shell})\frac{-i}{(\mathbf{v}-\mathbf{p})^2/4-\mu^2}+O(m^4v^4).
\end{aligned}\end{equation}
\end{widetext}
Since we also have $p_3^a=\frac{P_3+p_3}{2}=\frac{p_3}{2}$ and $p_3^b=\frac{P_3-p_3}{2}=\frac{-p_3}{2}$ we can simplify the prefactors of the dressed propagators:
\begin{equation}\begin{aligned}
    \frac{F(p^a_\mu)\pi}{f(p_\mu^a)}
    =\frac{eaV_0(2\pi)^4}{ p_3+ieaV_0}\\
\end{aligned}\end{equation}
and
\begin{equation}
    \begin{aligned}
        \frac{F(p^b_\mu)\pi}{f(p_\mu^b)}=\frac{eaV_0(2\pi)^4}{ -p_3+ieaV_0}
    \end{aligned}
\end{equation}
Therefore
\begin{widetext}
\begin{equation}\begin{aligned}
      h=\Gamma\left(\frac{P+p}{2}\right)\Gamma\left(\frac{P-p}{2}\right)/(-\Delta([P+p]/2)\Delta([P-p]/2))=
      \left(1+\frac{eaV_0(2\pi)^4}{ p_3+ieaV_0}\right)^2\end{aligned}\end{equation}
We perform the same operation on $G$:
\begin{equation}\begin{aligned}
    G=\int dp_0 \frac{i}{(p^a)^2-m^2+i\epsilon}\frac{i}{(p^b)^2-m^2+i\epsilon}&\left[\delta^4(p^a-u^a)+
       F(p^a_\mu) \delta^3(p^a_\mu-u^a_\mu)\frac{i}{(u^a)^2-m^2+i\epsilon}\right]\\
       \times&\left[\delta^4(p^b-u^b)+
       F(p^b_\mu) \delta^3(p^b_\mu-u^b_\mu)\frac{i}{(u^b)^2-m^2+i\epsilon}\right]\\
       =\frac{-4\pi i}{4m^2[\mathcal{E}-\frac{\mathbf{p}^2}{4m}]}\delta(E_\mathbf{p}-u^a_0)\delta(E_\mathbf{p}-u^b_0)&\left[\delta^3(\mathbf{p}/2-\mathbf{u}^a)+
       U(p_3)\delta^2(p_\nu/2-u^a_\nu)\frac{i}{(u^a)^2-m^2+i\epsilon}\right]\\
       \times&\left[\delta^3(-\mathbf{p}/2-\mathbf{u}^b)+
       U(p_3) \delta^2(-p_\nu/2-u^b_\nu)\frac{i}{(u^b)^2-m^2+i\epsilon}\right]\\
\end{aligned}\end{equation}
   
\end{widetext}
where we have used $f(p_\mu)=p_3\implies F(p_\mu)=  (1+\frac{ieaV_0}{2p_3})^{-1}eaV_0(2\pi)^3=\frac{eaV_0(2\pi)^3}{1+\frac{ieaV_0}{2p_3}}=U(p_3).$ We have also denoted $[\xi_1,\xi_2]=\xi_\nu$.

If we take $\lambda:=\frac{p_3}{eaV_0}\rightarrow 0$, hence the relative momentum $p_3^a-p_3^b$ in the tunneling direction to be much smaller than the product between the field-barrier coupling strength $e$ and barrier "area" $aV_0$, then the prefactor of each dressed propagator reduces to a complex constant $1-i2(2\pi)^3:=C$. If we also rewrite $U=\mathcal{U}(\lambda)=\frac{eaV_0(2\pi)^3\lambda}{\lambda+i/2}$, we see that under the strong barrier limit $\lambda\rightarrow0$, $\mathcal{U}(\lambda)\rightarrow 0$, and $G$ just reduces to the free resolvent $(\mathcal{E}-H_0)^{-1}$, hence demonstrating consistency with the free theory

Finally, we arrive at the 3D Lippmann-Schwinger equation, correct up to $O(m^4v^4)$ in the non relativistic limit:
\begin{widetext}
\begin{equation}
    L(\mathbf{p})=G(\mathbf{p})+\frac{\pi g^2}{m^2}\Lambda_-(p_3)\Lambda_+(p_3)\frac{1}{\mathcal{E}-H_0(\mathbf{p})}\int V(\mathbf{v}-\mathbf{p})L(\mathbf{v})d^3\mathbf{v}
\end{equation}
\end{widetext}
where $\Lambda_\pm(p_3)$ is (with the $2\pi$ factor adjusted) defined as before with the additional sign specification
\begin{equation}
    \Lambda_\pm(p_3)=1\pm\frac{-ieaV_0(2\pi)^4}{ \pm p_3+ieaV_0}.
\end{equation}
Since this is nothing but the one particle tunneling factor, $\Lambda_-(p_3)\Lambda_+(p_3)\frac{1}{\mathcal{E}-H_0(\mathbf{p})}$ is precisely the resolvent (in the fixed COM sector) of two independently tunneling particles. From this we conclude that (71) must be equivalent to a problem of category (1), with $W$ a delta function centered at $0$.

\section{Conclusion}
We have constructed a field theoretic, interactive two particle tunneling model using building blocks form a one particle tunneling theory, and shown that under approximations it permits closed solutions which may be readily extracted. We may now also confirm that such a model is consistent with known theoretical \cite{Zoellner2008FewBoson}, numerical \cite{Brugger2025}, and experimental \cite{Foelling2007SecondOrder} results by analyzing the leading order effect of interaction in a fully relativistic setting. On the other hand, we have also confirmed that our model is the relativistic generalization of the canonical and hard to solve class of Hamiltonians of form (1). These confirmations now justify and provide checks for the next steps towards explicitly obtaining non-perturbative solution via (40).

Without going into the non-perturbative regime for interparticle interaction, we have already been able to recover physical properties by the tunneling amplitude computed in section IV.A. The features of the "cross" structure, including the non-zero tunneling amplitude for near zero momentum, and the suppression of tunneling for in and out states with opposite momenta remains to be examined in more detail, whilst tuning the parameters should reveal even more diverse structures. Whether they are physical or mathematical artifacts is also of question, and if they may be observed experimentally, the current model may serve as a good testing ground for understanding new quantum phenomena. 

More fundamentally, our results demonstrate that a fully relativistic two-body tunneling framework can reproduce nontrivial correlated effects without invoking phenomenological potentials, thereby bridging few-body quantum mechanics and field-theoretic formulations. Unlike conventional non-relativistic tunneling models that rely on effective potentials or separable ansätze, the present construction arises directly from the underlying propagator structure of a relativistic field theory. This ensures that particle correlations, retardation effects, and virtual excitations are all treated on an equal footing, rather than being imposed by hand. In this sense, the model serves as a minimal yet self-consistent realization of interacting tunneling in which both the single-particle and two-particle amplitudes emerge from the same dynamical kernel.

\begin{acknowledgments}
I express my deepest gratitude to Professor LIANG Haozhao at the University of Tokyo for his tireless mentorship, to whom I also owe the discovery of this problem. I also thank Gabriel Mayor and Hadrien Deo for allowing me to join their bachelor’s thesis project, and Chisato Sakaki for reviewing my methods. I am further grateful to Professor Tsuganuma at Kyoto University, Professor Yamazaki at the University of Tokyo, Professor Simenel at the Australian National University, and Professors Noah and Mack at Reed College, for their valuable discussions and feedback.
\end{acknowledgments}

\bibliographystyle{unsrtnat}
\bibliography{refs}

\appendix
\section{Deriving the Tunneling Propagator}
Denote the ${S}$-matrix for $\mathcal{L}_{tun}$ as $S_{tun}$. Since (3) generates only linear graphs, amputation only removes the in and out propagators for any graph. Therefore, to recover $D_\psi(p,q)$, we first obtain the off-shell version of $\braket{p|S_{tun}|k}$, which we denote $\mathcal{M}(p,k)$, which is just the sum of the amputated diagrams without on shell condition $\mathcal{M}^{(0)}(p,k)+\mathcal{M}^{(1)}(p,k)+\cdots$.  We then reattach the bare $\psi$ in and out propagators $i\Delta_\psi(p)=\frac{i}{p^2-m^2+i\epsilon}$. This gives
\begin{equation}\begin{aligned}
    D_\psi(p,q)=i\Delta_\psi(p)\mathcal{M}(p,k)i\Delta_\psi(q).
\end{aligned}\end{equation}
The trivial term denoted $\braket{p|S_{tun}|k}^{(0)}$ in \cite{Zielinski2024b} becomes the free propagator $(2\pi)^4\delta^4(p-k)i\Delta_\psi(p)$.  The next term was computed without invoking the on-shell condition so  $\braket{p|S_{tun}|k}^{(1)} =\mathcal{M}^{(1)}(p,k)$. The rest of the diagrams are still resummed as \cite{Zielinski2024b} does in its equations (15) to (18), with the only difference being that the on-shell condition is removed, so we do not have $
p_{3} = \sqrt{p_{0}^{2} -p_{1}^{2} - p_{2}^{2} - m^{2}}$ and do not rewrite the pole in (15) of \cite{Zielinski2024b} as $p_3$. Therefore the transition of the recursive relation (16) of \cite{Zielinski2024b} from $\braket{p| S_{tun}|k}$ to $\mathcal{M}(p,k)$ is simply 
\begin{align*}
\langle k|S_{tun}|p\rangle^{(2)}
&=
\frac{(-\,i e\,a\,V_0)^2}{2\,p_3}\,(2\pi)^3
\prod_{\mu=0}^{2}\delta\bigl(p_\mu - k_\mu\bigr)\\
&=
-\frac{i e\,a\,V_0}{2\,p_3}\;\langle k|S_{tun}|p\rangle^{(1)}\,
\\
&\hspace{2.5cm}\Bigg\downarrow\text{off shell}\\
\mathcal{M}^{(2)}(p,k)
&=
\frac{(-\,i e\,a\,V_0)^2}{2\, \sqrt{p_{0}^{2} -p_{1}^{2} - p_{2}^{2} - m^{2}}}\,(2\pi)^3
\prod_{\mu=0}^{2}\delta\bigl(p_\mu - k_\mu\bigr)\\
&=
-\frac{i e\,a\,V_0}{2\, \sqrt{p_{0}^{2} -p_{1}^{2} - p_{2}^{2} - m^{2}}}\;\mathcal{M}^{(1)}(p,k).
\\
\end{align*}

\section{Simplification of the Bethe-Salpeter Equation}

The BS equation for the unamputated off shell S Matrix reads:
\begin{widetext}
\begin{equation}\begin{aligned}
    L(p^a,p^b;u^a,u^b)=&D_\psi(p^a,u^a)D_\psi(p^b,u^b)+g^2\int D_\psi(p^a,k^a)D_\psi(p^b,k^b)D_{\phi}(q) L(v^a,v^b;u^a,u^b)\\
   & \times\delta^2(k^a+q-v^a)\delta^2(k^b-q-v^b) d^4k^a d^4k^bd^4qd^4v^ad^4v^b
\end{aligned}\end{equation}
second delta function eliminates q integral and gives $q=k^b-v^b$ This gives the integral
\begin{equation}\begin{aligned}
    I=\int D_\psi(p^a,k^a)D_\psi(p^b,k^b)D_{\phi}(k^b-v^b) L(v^a,v^b;u^a,u^b)\delta^2(k^a+k^b-v^a-v^b)d^4k^a d^4k^bd^4v^ad^4v^b.
\end{aligned}\end{equation}
Change to COM coordinates for all three variables
\begin{equation}\begin{aligned}
    \xi^a+\xi^b&=\Xi\\
    \xi^a-\xi^b&=\xi\\
    d^4 \xi^a d^4\xi^b& \rightarrow d^4\Xi d^4\xi
\end{aligned}\end{equation}
And take frame such that $\mathbf{P}=0$, hence $P=(E,\mathbf{0})$. This does not apply for $U$ since translational symmetry is broken by the delta barrier.
\begin{equation}\begin{aligned}
    I&=\int D_\psi(p^a,k^a)D_\psi(p^b,k^b)D_{\phi}\left(\frac{K-k-(V-v)}{2}\right) L(v^a,v^b;u^a,u^b)\delta^2(K-V)\,d^4K \,d^4k\,d^4V\,d^4v\\
    &=\int D_\psi(p^a,k^a)D_\psi(p^b,k^b)D_{\phi}\left(\frac{v-k}{2}\right) L(K,v;U,u)\,d^4K \,d^4k\,d^4v
\end{aligned}\end{equation}
Now we revert the $k$ coordinates back $ d^4K d^4k  \rightarrow d^4 k^a d^4 k^b$:

\begin{equation}\begin{aligned}
  I  =\int D_\psi(p^a,k^a)D_\psi(p^b,k^b)D_{\phi}\left(\frac{v-(k^a-k^b)}{2}\right) L\left(\frac{(k^a+k^b)+v}{2},\frac{(k^a+k^b)-v}{2};u^a,u^b\right)\,d^4K \,d^4k\,d^4v
\end{aligned}\end{equation}
Recall that
\begin{equation}\begin{aligned}
       D_\psi(p^a,k^a)=\frac{i}{(p^a)^2-m^2+i\epsilon}\delta^4(p^a-k^a)+
       F(p_\mu^a) \frac{i}{(p^a)^2-m^2+i\epsilon}\delta^3(p^a_\mu-k^a_\mu)\frac{i}{(k^a)^2-m^2+i\epsilon}
\end{aligned}\end{equation}
then performing the integration over $k$ will collapse it as $k\rightarrow p$. This is trivial for $k_\mu$: after performing this integral first we have for some arbitrary but reasonably well behaved distribution $g$
\begin{equation}\begin{aligned}
   &\int D_\psi(p^a,k^a)g(k_\mu,k_3)dk^a_\mu\\
   &=\left[\frac{i}{(p^a)^2-m^2+i\epsilon}\delta(p^a_3-k^a_3)+F(p^a_\mu) \frac{i}{(p^a)^2-m^2+i\epsilon}\frac{i}{(p_\mu^a)^2-(k_3^a)^2-m^2+i\epsilon}\right]g(p_\mu^a,k_3^a).
\end{aligned}\end{equation}
Then, integrating over $k_3^a$ is trivial for the first term. For the second term we have to employ contour integration. The poles are located at $k=\pm\left(\sqrt{p_\mu^2-m^2}+i\epsilon\right)$ with $\epsilon>0$. We pick up the positive pole for $k^a$ and get 
\begin{equation}\begin{aligned}
    \int D_\psi(p^a,k^a)g(k_\mu^a,k^a_1)dk_\mu^a dk_3^a&=\left[\frac{i}{(p^a)^2-m^2+i\epsilon}+F(p^a_\mu)\frac{i}{(p^a)^2-m^2+i\epsilon}\frac{\pi}{f(p^a_\mu)}\right]g(p^a_\mu,f(p^a_\mu))\\
    &=\Gamma(p^a)g(p^a_\mu,f(p_\mu^a)).
\end{aligned}\end{equation}
If for $k^b$ we pick up the negative pole instead, we get 
\begin{equation}\begin{aligned}
    \int D_\psi(p^b,k^b)g(k_\mu^b,k^b_1)dk_\mu^b dk_1^b&=\left[\frac{i}{(p^b)^2-m^2+i\epsilon}-F(p_\mu^b)\frac{i}{(p^b)^2-m^2+i\epsilon}\frac{\pi}{f(p_\mu^b)}\right]g(p_\mu^b,-f(p_\mu^b))\\
    &=\Gamma(p^b)g(p_\mu^b,-f(p_\mu^b)).
\end{aligned}\end{equation}
we have collapsed $k^a,k^b$ such that $k^a,k^b\rightarrow[p_\mu^a,f(p_\mu^a)],[p_\mu^b,-f(p_\mu^b)]$. For simplicity let us define $[p_\mu^a,f(p_\mu^a)]-[p_\mu^b,-f(p_\mu^b)]=p'$. Also, $K=[p_\mu^a,f(p_\mu^a)]+[p_\mu^b,-f(p_\mu^b)]=[0,f(p_\mu^a)-f(p_\mu^b)]:=P'$

Therefore after integrating $k^a,k^b$, the wider integral becomes
\begin{equation}\begin{aligned}
    I&=\int \Gamma(p^a)\Gamma(p^b)D_{\phi}\left(\frac{v-p'}{2}\right) L(P',v;U,u)\,d^4v\\
    &= \Gamma(p^a)\Gamma(p^b)\int \frac{i}{(v_\mu-p_\mu)^2/4-(v_3-[f(p_\mu^a)+f(p_\mu^b)])^2/4-\mu^2+i\epsilon}L(P',v;U,u)d^3v_\mu dv_3,
\end{aligned}\end{equation}

and the BS equation reads (inhomogenous)
\begin{equation}\begin{aligned}
    L(P,p;U,u)=&D_\psi\left(\frac{P+p}{2},\frac{U+u}{2}\right)D_\psi\left(\frac{P-p}{2},\frac{U-u}{2}\right)\\
    +&\Gamma\left(\frac{P+p}{2}\right)\Gamma\left(\frac{P-p}{2}\right)g^2\int \frac{i}{(v-p')^2/4-\mu^2+i\epsilon}L(P',v;U,u)d^4v.
\end{aligned}\end{equation}
\end{widetext}
\section{Positive energy approximation in the NR limit}
In our case we have 
\begin{align}
    g(p_0)=\int &\frac{-i}{(v_1-p'_1(p_0))^2/4+\mu^2-i\epsilon} L(v_1)dv_1 
\end{align}
We use again the parameter defined in IV.B. Denote the soft and hard momentum as
\begin{align}
p_0^{(-)}&:=-E+\sqrt{p_1^2+4m^2}=O(mv^2)\\
p_0^{(+)} &:= E + \sqrt{p_1^2 + 4m^2}
= 4m + O(m v^2),
\end{align}
which gives 
\[
p'_1(p_0^{(-)})=O(mv),\qquad p'_1(p_0^{(+)})=O(m).
\]
Crucially, we assume that $L(v_1)$ dominates for $v_1=O(mv)$, hence $|v_1|=O(mv)$ on the support of $L(v_1)$. Then, the soft case gives
\begin{align}
g(p_0^{(-)})
=
O\!\left(\frac{\int |L(v_1)|\,dv_1}{m^2v^2+\mu^2}\right).
\end{align}
For the hard case:
\begin{align}
    \left| v_1 - p_1'\!\left(p_0^{(+)}\right) \right|
\ge
\left| p_1'\!\left(p_0^{(+)}\right) \right|
-
|v_1|
\\=
O(m) - O(mv)
=
O(m),
\end{align}
which gives 
\begin{align}
g(p_0^{(+)})
=
O\!\left(
\frac{ \int |L(v_1)|\, dv_1}{m^2+\mu^2}
\right).
\end{align}
Hence
\begin{align}
    \frac{g(p_0^{(+)})}{g(p_0^{(-)})}
=
O\!\left(\frac{m^2v^2+\mu^2}{m^2+\mu^2}\right).
\end{align}
However, since we assume $\mu\ll m$, we get
\begin{align}
    \frac{g(p_0^{(+)})}{g(p_0^{(-)})}
=
O\!\left(v^2\right).
\end{align}
Hence neglecting the $g(p_0^{(+)})$ in (21) in the NR regime is a controlled approximation.
\end{document}